\documentclass[prb,twocolumn]{revtex4}

\usepackage{times}
\usepackage{amsmath}
\usepackage{amssymb}
\usepackage{psfrag}
\usepackage{pstricks}
\usepackage{graphicx}
\usepackage{dcolumn}
\usepackage{bm}
\usepackage[hypertex]{hyperref}

\def\be{\begin{equation}}
\def\ee{\end{equation}}

\begin{document}

\title{Odd-Even Crossover in a non-Abelian
$\nu=5/2$ Interferometer}

\author{Waheb Bishara}
\affiliation{ Department of Physics, California Institute of
  Technology, MC 256-80 Pasadena, CA 91125}
\author{Chetan Nayak}
\affiliation{Microsoft Research, Station Q, CNSI Building,
University of California, Santa Barbara, CA 93106-4030}
\affiliation{Department of Physics,
University of California, Santa Barbara, CA 93106}

\begin{abstract}
We compute the backscattered current in a double point-contact geometry of a Quantum Hall system at filling fraction $\nu=5/2$
as a function of bias voltage in the weak backscattering regime.
We assume that the system is in the universality class
of either the Pfaffian or anti-Pfaffian state. When the
number of charge $e/4$ quasiparticles in the interferometer
is odd, there is no interference pattern with period $4\Phi_0$
at temperatures and source-drain voltages high enough that
the coupling between bulk quasiparticles and the edge can be neglected.
However, the coupling between a bulk charge $e/4$ quasiparticle
and the edge causes it to be effectively absorbed by the edge
at low temperatures and voltages.
Consequently, even with an odd number of $e/4$ quasiparticles
in the interferometer, an interference pattern with period $4\Phi_0$
appears at low bias voltages and temperatures,
as if there were an even number of
quasiparticles in the interferometer. We relate this
problem to that of a semi-infinite Ising model with
a boundary magnetic field. Using the methods
of perturbed boundary conformal field theory,
we give an exact expression for this crossover
of the interferometer as a function of bias voltage.
Finally, we comment on the possible relevance of
our results to recent interference experiments.
\end{abstract}
\maketitle

\section{Introduction}

A two point-contact interferometer
\cite{Chamon97,Fradkin98,Camino05}
is potentially a valuable probe of the topological
properties of quantum Hall states. If the observed
state at $\nu=5/2$ \cite{Willett87,Eisenstein02,Xia04}
were non-Abelian \cite{Moore91,Nayak96c,Ivanov01},
there would be a very dramatic signature in
transport through a two point-contact interferometer
\cite{Bonderson06a,Stern06,Overbosch07,Bishara08}.
If there is an even number of charge $e/4$
quasiparticles in the interferometer, then Aharonov-Bohm
oscillations of the current are observed as the area of the loop is varied, due to the interference between the two possible
tunneling paths for current-carrying charge $e/4$ quasiparticles.
If there is an odd number of quasiparticles in the
loop, then these Aharonov-Bohm oscillations are not
observed as a result of the non-Abelian braiding
of the current-carrying $e/4$ quasiparticles with those in
the bulk. (However, Aharonov-Bohm oscillations
with twice the period will still be observed
due to the current carried by charge $e/2$ quasiparticles \cite{Bishara09}.)
A recent experiment \cite{Willett09} may have observed
this predicted effect.

In this experiment, a side gate is used to vary the area
of the quantum Hall droplet in the interferometer.
The current oscillates as the area is varied.
However, at certain values of the side gate voltage,
the interference pattern changes dramatically.
According to the non-Abelian interferometry interpretation,
such a change occurs when the area is varied
beyond a point at which one of the quasiparticles leaves 
the interference loop. Then the $e/4$ quasiparticle
number parity in the interference loop changes, leading
to a striking change in the interference pattern.
Close to a transition point in the $e/4$ quasiparticle
number parity, a quasiparticle comes close to the edge of the
quantum Hall droplet and begins to interact with the edge
excitations. The leading coupling of the $e/4$ quasiparticle
to the edge is through the (resonant) tunneling of Majorana
fermions from the edge to the zero mode on the
$e/4$ quasiparticle. This coupling makes it possible for $e/4$
Aharonov-Bohm oscillations to be seen even when there
is an odd number of quasiparticles in the interference loop.
At an intuitive level, this can be understood in the following way.
For odd quasiparticle number, a topological qubit straddles one of the point contacts
and records when an $e/4$ quasiparticle takes that path; consequently
the two paths do not interfere and Aharonov-Bohm oscillations are not
seen. This qubit is flipped when a Majorana fermion tunnels from the edge
to a bulk zero mode in the interference loop, thereby erasing the record
and allowing quantum interference. Over longer time scales,
the topological qubit flips so many times that it can no longer carry any information.
This eventually leads, at low energies and long time scales, to
the absorption of the zero mode by the edge and, therefore,
to the effective removal of this quasiparticle from the interference
loop, as far as its non-Abelian braiding properties are
concerned. Thus, {\it every} bulk quasiparticle
will appear to be effectively absorbed by the edge
if the interferometer is probed at
sufficiently low voltages and temperatures -- but
`sufficiently low' will be exponentially small
in the distance of the quasiparticle from the edge,
as we will see. Thus, the effect of bulk-edge coupling
will only be apparent when the edge is close to a bulk $e/4$ quasiparticle.
In this paper, we analyze this coupling in detail,
as it effects the behavior of a two point-contact interferometer,
with possible relevance to the transition regions of the
experiments of Refs \onlinecite{Willett09}.

In Ref. \onlinecite{Fendley09}, the coupling of a bulk
$e/4$ quasiparticle to the edge was formulated in
terms of perturbed boundary conformal field theory.
It was shown that this problem could be mapped to
a semi-infinite Ising model in a boundary magnetic field.
As we discuss below, the absence of $e/4$ quasiparticle
interference for an odd number of bulk quasiparticles
corresponds to the vanishing of the one-point function
$\langle\sigma(x)\rangle=0$ when the
boundary magnetic field vanishes, while the appearance
of $e/4$ quasiparticle interference for an even number of
bulk quasiparticles corresponds to
$\langle\sigma(x)\rangle=x^{-1/8}$ when the
boundary magnetic field is infinite ($x$ is the distance to
the boundary of the Ising model which is assumed, for simplicity,
to be the $y$-axis). For finite boundary magnetic field,
the boundary conditions of the Ising model cross over
from free to fixed, which corresponds to the absorption
of a bulk quasiparticle. Following the derivation of
of the exact crossover function for the magnetization
by Chatterjee and Zamolodchikov \cite{Chatterjee94}
(and of the full boundary state by Chatterjee \cite{Chatterjee95}),
we compute the current through the interferometer
to lowest order in the backscattering at the point contacts,
but treating the bulk-edge coupling {\it exactly}.
Our results agree with lowest order perturbation
theory in the bulk-edge coupling \cite{Overbosch07}
and numerical solution of a lattice model \cite{Rosenow07b}.
When the point contacts are close together compared to
${v_n}/{e^*}V$, where $v_n$ is the Majorana fermion edge
velocity, ${e^*}=e/4$, and $V$ is the source-drain voltage,
the current-voltage  relation takes a particularly simple form. 
When there is an even number of quasiparticles in the bulk, one of
which is close to the edge, an interesting non-equilibrium
problem presents itself: suppose the internal topological
state of the bulk quasiparticles is fixed to an initial value;
what is its subsequent time evolution. This is considered
elsewhere.

\section{Model}

We now set up the calculation of the backscattered current
in a two point-contact interferometer to lowest order. The Pfaffian and anti-Pfaffian cases are conceptually similar, so we focus on the Pfaffian for the sake of concreteness.
The edge theory of the Pfaffian state has a chiral bosonic charge mode and a chiral neutral Majorana
mode\cite{Milovanovic96,Bena06,Fendley06,Fendley07a}
\be
\label{eqn:Pf-edge}
{\cal L}^{R}_{\rm Pf}(\psi,\phi)=\frac{2}{4\pi}\partial_x\phi\left(\partial_t+v_c\partial_x\right)\phi + i\psi\left(\partial_t+iv_n\partial_x\right)\psi
\ee
Both modes propagate to the right (the left-moving version of
this action, ${\cal L}^{R}_{\rm Pf}$, has time-derivative
terms with opposite sign), but will have different velocities in general.
The velocities of the charged and neutral modes
are $v_c$ and $v_n$, respectively. Recent numerical calculations
and experiments indicate that ${v_c}\sim 10^5$ m/s, while ${v_n}\sim {v_c}/10$
(see, for instance, Ref. \onlinecite{Wan06,Zhang09}).
The electron operator
and $e/4$ quasiparticle operators are, respectively,
$\Phi_{el}=\psi e^{i\sqrt{2} \phi}$ and
$\Phi_{1/4}=\sigma e^{i\phi/2\sqrt{2}}$,
where $\sigma$ is the Ising spin field of the Majorana fermion
theory\cite{Fendley07a}.

\begin{figure}[b!]
\begin{center}
  \includegraphics[width=3.25in]{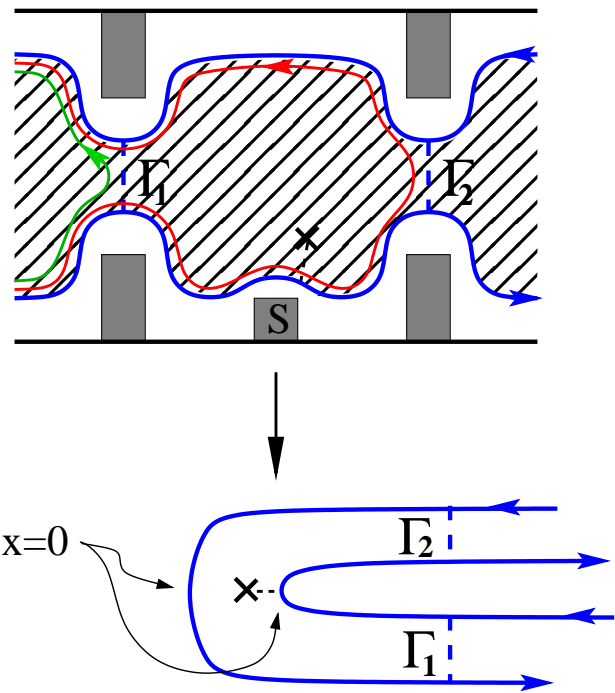}
  \caption{A double point-contact interferometer. Edge quasiparticles tunnel at two point-contacts with amplitudes $\Gamma_1$
and $\Gamma_2$, respectively. The interferometry area is changed by applying a voltage to gate $S$. A bulk quasiparticle
is coupled to the bottom edge by Majorana fermion
tunneling. This setup can be reformulated as two
semi-infinite non-chiral edges or, equivalently,  two
semi-infinite Ising models. One Ising model has fixed boundary
condition; the other had free boundary condition and a boundary
magnetic field.}
  \label{fig:interfero}
\end{center}
\end{figure}

In the interferometer geometry depicted in Fig. \ref{fig:interfero},
the edge modes are oppositely directed on the bottom and top
edges, which we done by the subscripts $1,2$. The two point
contacts are at $x_a$, $x_b$,
and the corresponding $e/4$ quasiparticle backscattering
amplitudes are $\Gamma_a$, $\Gamma_b$.
Experimental values \cite{Willett09,Zhang09}
of $|{x_a}-{x_b}|$ can range from approximately
$1 \mu$m to $5 \mu$m, while
${|\Gamma_{a,b}|^2}\sim 0.1$.
In the absence of backscattering
at the point contacts or bulk-edge coupling,
the action of the device is:
\be
\label{eqn:Pf-two-edges}
{S_0} = \int dt \int dx \,\left(
{\cal L}^{R}_{\rm Pf}({\psi_1},{\phi_1}) + 
{\cal L}^{L}_{\rm Pf}({\psi_2},{\phi_2})\right)
\ee

When inter-edge backscattering is weak, we expect the amplitude
$\Gamma$ for charge-$e/4$ to be transferred from one edge to the other to be larger than for higher charges
$ne/4$ \cite{Bishara09}. It is also the most relevant
backscattering operator in the Renormalization Group sense
\cite{Fendley06,Fendley07a},
so we will focus on it. Since it is relevant, its effective value grows
as the temperature is decreased, eventually leaving the weak
backscattering regime. We assume that the temperature or voltage is
high enough that the system is still in the weak inter-edge
backscattering regime and a perturbative calculation is valid,
but still much lower than the bulk energy gap.
Following Refs. \onlinecite{Chamon97,Bishara08},
inter-edge backscattering
leads to a term of the form
\begin{multline}
\label{eqn:tunneling-action}
S_\text{backscatt} = \int dt\,\,\bigl(
{\Gamma_a} e^{-i{\omega^{}_J}t}\,\, {T_a}(t)
 \:+\: \text{c.c.} \: +\\
{\Gamma_b} \,e^{2\pi i(\frac{\Phi}{4\Phi_0}-\frac{{n_q}}{8}
+ \frac{n_\psi}{2})}
\: e^{-i{\omega^{}_J}t}\,\,
{T_b}(t)
\:+\: \text{c.c.}\bigr)
\end{multline}
where
\begin{equation}
\label{eqn:backscat-op}
T_{a}(t) = {\sigma_1}(x_{a},t)\,{\sigma_2}(x_{a},t)\,\,
e^{\frac{i}{\sqrt{8}}\left({\phi_1}(x_{a},t)-{\phi_2}(x_{a},t)\right)}
\end{equation}
and similarly for ${T_b}(t)$. The Josephson frequency
for a charge $e/4$ quasiparticle with voltage $V$ applied
between the bottom and top edges is
$\omega^{}_J={e^*}V=\frac{eV}{4}$ (in units in which $\hbar=1$).
The difference in the magnetic fluxes enclosed by the two
trajectories around the interferometer is $\Phi$. We have chosen a gauge in which the vector potential is concentrated at the second point contact so that $\Phi$ enters only through the second term above.
$n_q$ is the total electrical charge of the bulk quasiparticles,
in units of $e/4$; ${n_\psi}=0,1$ is the Majorana fermion
number in the interference loop, modulo $2$. The $n_q$ and $n_\psi$
terms in $\Gamma_b$ account for the diagonal (in the
fermion number basis) effects of quasiparticle statistics.
The product of right- and left-moving spin fields in
(\ref{eqn:backscat-op}) must be handled with some care
to account for the fact that two charge $e/4$ quasiparticles
(one on each edge) can fuse in two different ways.
The effect of the non-Abelian braiding statistics
of the bulk quasiparticles enters in this way through the
precise definition of $T_{a,b}$.
Fortunately, this can be handled in a simple way
in a calculation to lowest-order in the backscattering operator,
as we will see in the next section.

We now consider the coupling between
a bulk quasiparticle and the edge. Suppose that one
of the bulk quasiparticles is close to the bottom edge, at
$x=x_0$ with ${x_a}<{x_0}<{x_b}$, as depicted
in Fig. \ref{fig:interfero}. Each bulk $e/4$ quasiparticle has a
Majorana fermion zero mode \cite{Read96,Read00};
we will denote
the zero mode associated with the $e/4$ quasiparticle
close to the edge by $\psi_0$. Then, the leading coupling between
the edge and this quasiparticle is of the form:
\begin{equation}
\label{eqn:bulk-edge-coupling}
S_\text{bulk-edge} = \int dt\,\left(
{\psi_0}{\partial_t}{\psi_0} + 
2ih\,{\psi_0}\, {\psi_1}\!({x_0})\right)
\end{equation}
Here, $2h$ is the amplitude for a Majorana fermion to
tunnel from the edge to the zero mode $\psi_0$.

Thus, the total action for a two-point contact interferometer
with one or more quasiparticles in the interference loop,
one of which is close to the bottom edge, is of the form
\begin{equation}
\label{eqn:total-interfero-action}
S = {S_0} + S_\text{backscattering} + S_\text{bulk-edge}
\end{equation}
However, this description is, at the moment, incomplete
because we have not precisely defined the product
of Ising spin fields in $S_\text{backscattering}$.
We will do this in the next section, but first
we will give the appropriate Kubo formulae for
current through the interferometer.

The current operator can be found from the commutator
of the backscattering Hamiltonian and the charge on one edge:
\begin{multline}
\label{eqn:current-op}
I(t)=\frac{ie}{4}\left(\Gamma_a e^{-i\omega^{}_J t}\,{T_a}(t) -
\text{h.c.}\right)\\
+ \frac{ie}{4}\left(
{\Gamma_b} \,e^{2\pi i(\frac{\Phi}{4\Phi_0}-\frac{{n_q}}{8}
+ \frac{n_\psi}{2})}
\: e^{-i{\omega^{}_J}t}\,
{T_b}(t) - \text{h.c.}\right)
\end{multline}
To lowest order in perturbation theory, the backscattered
current is found to be:
\be\label{Current}
\langle I(t) \rangle = -i \int_{-\infty}^t dt' \, \langle 0 |
[I(t),H_\text{backscatt}(t')]|0\rangle
\ee
In principle, the current must be computed using a non-equilibrium
technique, such as the Schwinger-Keldysh method, when
the voltage is finite. However, at first order in the backscattering
operators, there is no difference between the Schwinger-Keldysh expression
and (\ref{Current}).

At this order, the current
naturally breaks into the sum of three terms
$I = I_{a} + I_{b} + I_{\rm int}$ where
\begin{multline}
I_{a,b} = \frac{e}{4}
\left|\Gamma_{a,b}\right|^2 \,
\int_{-\infty}^0 \!\! dt \, e^{i{\omega^{}_J}t} \bigl(
\langle {T_{a,b}^{}}(0)  {T_{a,b}^\dagger}(t)\rangle\\
 -  \langle{T_{a,b}^\dagger}(0) {T_{a,b}^{}}(t) \rangle
\bigr)
\end{multline}
are the backscattered currents for each point contact
independently and, following Chamon {\it et al.}\cite{Chamon97},
we write the interference term in the form:
\begin{align}
\label{eqn:interference-term}
I_{\rm int} &= \frac{e}{4}\,
\Gamma_a \tilde{\Gamma}_b^* \,
\int_{-\infty}^0 \!\! dt \, e^{i{\omega^{}_J} t} \bigl(
\langle {T_a^{}}(0)  {T_b^\dagger}(t)\rangle
-  \langle {T_b^\dagger}(0) {T_a^{}}(t) \rangle
\bigr)\cr
&  \hskip 0.7 cm + \: \text{c. c.}\cr
&=  \frac{e}{4}\cdot 2\,\text{Re}\Bigl(\Gamma_a {\tilde{\Gamma}_b^*}
  \bigl[\tilde{P}({\omega^{}_J})
-\tilde{P}(-{\omega^{}_J})\bigr]\Bigr)
\end{align}
where $\tilde{\Gamma}_b=
{\Gamma_b} \,e^{2\pi i(\frac{\Phi}{4\Phi_0}-\frac{{n_q}}{8}
+ \frac{n_\psi}{2})}
$.
and the imaginary part of the response function is
\begin{equation}
\tilde{P}({\omega^{}_J}) = \int_{-\infty}^\infty \!\! dt\, e^{i{\omega^{}_J} t}\,
\langle {T_a^{}}(0)  {T_b^\dagger}(t) \rangle
\end{equation}
$I_{\rm int}$ is due to interference between the process in which
a quasiparticle tunnels between the two edges
at $x_a$ and the process in which it continues to $x_a$ and tunnels there. As a result, $I_{\rm int}$ depends on the magnetic flux and
the number of bulk quasiparticles between the two point contacts;
it reflects the non-Abelian statistics of quasiparticles. This
is implemented through the precise definition of the product
of Ising spin fields which appears in the backscattering operators,
to which we turn in the next section.

\section{Backscattering Operators and Interference}

We now review the a few essential points in
the discussion of inter-edge backscattering
in Refs. \onlinecite{Fendley06,Fendley07a}.
In the chiral Ising model, a pair of $\sigma$s can fuse
to either $1$ or $\psi$
(or any linear combination of the two). Consequently,
when we consider the correlation function of a string of $2n$
(chiral) $\sigma$ fields at a single edge, there is not
a unique answer but, instead, a vector space of
$2^{n-1}$ {\it conformal blocks} which are defined by
specifying the fusion channels of the $\sigma$s (e.g. by
dividing them arbitrarily into $n$ pairs and specifying
how each pair fuses; different pairings lead to different
bases in the vector space).

When a charge $e/4$ quasiparticle backscatters from one
edge to another, a pair of $\sigma$ quasiparticles is created,
one in the non-Abelian sector of each edge
(recall that a $\sigma$, which is the non-Abelian
part of an $e/4$ quasiparticle, is its own anti-particle).
When there is only a single point contact and all bulk
quasiparticles are far from the point contact, we can take this
pair of $\sigma$s to fuse to $1$ since the
backscattering process is a very small motion of
a quasiparticle which does not involve any braiding
and, therefore, does not create a $\psi$. An alternative
way to understand this is to note that
one can choose a gauge in which the
non-Abelian gauge field due to bulk quasiparticles vanishes
at the point contact. In this way,
we can give a precise meaning to operators such as $T_{a,b}$.
However, when we compute perturbatively in the backscattering,
we would like to know how successive $\sigma$ fields
on the {\it same} edge fuse. Fortunately, the condition
that the pair of $\sigma$s which is created on opposite edges
by a backscattering event can be
converted (using a feature of anyon systems called
the $F$-matrix) into a condition on the fusion channels
of successive $\sigma$ fields on the same edge.
This leads, according to the arguments of
Refs. \onlinecite{Fendley06,Fendley07a}, to a mapping of
the single point contact problem to a Kondo-esque
impurity problem.

When there are two point contacts, the quasiparticle
history associated with a backscattering process at
one of the point contacts must necessarily wind around the
bulk quasiparticles in the interferometer. Equivalently,
the non-Abelian gauge field due to bulk quasiparticles
is non-vanishing at one of the point contacts or along
one of the edges between the two point contacts; the
simplest gauge is one in which the gauge field is concentrated
at one of the point contacts, say contact $b$. If there is an
even number of quasiparticles in the interferometer,
their effect can be encapsulated in an extra phase ${n_\psi} \pi$
(dependent on the overall parity of the topological qubits
in the interference loop) which we have absorbed
into $\Gamma_b$. However, as shown in
Ref. \onlinecite{Bishara08}, if there is an odd
number of quasiparticles in the loop, then the pair
of $\sigma$s which is created by $T_{b}$ must fuse
to $\psi$ instead of $1$. This makes no difference as far
as local properties of that point contact are concerned.
(In fact, in the single point-contact problem, we could
have taken each backscattered pair to fuse to $\psi$ instead of
to $1$, which would correspond to a non-standard
gauge choice. This would have no effect on any
physical property and would lead to the same Kond-esque model.)

However, if we consider the interference between
the backscattering processes due to $T_{a}$ and $T_{b}$
when there is an odd number of quasiparticles in the interference
loop, it is significant that the pair of $\sigma$s created by the
former fuse to $1$ but those created by the latter fuse to $\psi$.
If we consider the current to lowest order in the backscattering,
the interference term (\ref{eqn:interference-term})
contains the expression
\begin{align*}
\langle {T_a^{}}(0)  {T_b^\dagger}(t)\rangle &= 
\langle {\left[{\sigma_1}({x_a},0)\,{\sigma_2}({x_a},0)\right]_1}\,\,
{\left[{\sigma_1}({x_b},t)\,{\sigma_2}({x_b},t)\right]_\psi}
\rangle\cr
&= \langle {\left[{\sigma_1}({x_a},0)\,{\sigma_1}({x_b},t)\right]_\psi}\rangle\,\,
 \langle{\left[{\sigma_2}({x_a},0)\,{\sigma_2}({x_b},t)\right]_1}
\rangle\cr
& \hskip 0.5 cm + \: 1 \leftrightarrow 2
\end{align*}
Here, we have used square brackets to denote the fusion
channels of pairs of $\sigma$ fields. The correlation function
in the first line factorizes, as shown on the second line,
because we are perturbing around the limit in which the
edges are decoupled.
This expression vanishes because
$\langle{\left[{\sigma_2}({x_a},0)\,{\sigma_2}({x_b},t)\right]_\psi}\rangle$
vanishes by fermion number parity conservation. However,
when a bulk quasiparticle is coupled to the edge according to
(\ref{eqn:bulk-edge-coupling}), fermion number parity is no
longer conserved. Thus, this correlation function need not
vanish, and an interference term can be present even for
odd quasiparticle numbers \cite{Overbosch07,Rosenow07b}.
For instance, to lowest order in the tunneling amplitude $h$
in (\ref{eqn:bulk-edge-coupling}),
$\langle {T_a^{}}(0)  {T_b^\dagger}(t)\rangle$
will contain a non-vanishing contribution of the form \cite{Overbosch07}
$$
h\langle \psi_0 \rangle\,
\langle {\psi_1}({x_0},t')\, {\left[{\sigma_1}({x_a},0)
{\sigma_1}({x_b},t)\right]_\psi}\rangle\,
\langle {\left[{\sigma_2}({x_a},0)\,{\sigma_2}({x_b},t)\right]_1}\rangle\,
$$
As discussed in Ref. \onlinecite{Overbosch07}, this leads to
a non-vanishing interference term for odd quasiparticle
numbers with different scaling properties (as a function
of $T,V$) than for even quasiparticle numbers. In the
next section, we show how this interference
term can be computed exactly in $h$ (but still to lowest
order in ${\Gamma_a}\tilde{\Gamma}_b^*$).

\section{Mapping to the Ising Model with a Boundary}
\label{sec:mapping}

When there is an even number of $e/4$ quasiparticles 
in the bulk of a Pfaffian or anti-Pfaffian droplet,
the Majorana fermions have anti-periodic boundary conditions.
When there is an odd number, the Majorana fermions
have periodic boundary conditions. This can be understood
in terms of the classical critical $2D$ Ising model in the
following way\cite{Fendley09}. The droplet is `squashed'
down so that the bottom and top edges become the
right- and left-moving modes of the Ising model. The bulk is
forgotten about, except insofar as it affects the boundary
conditions at the two ends of the droplet, where right-moving
modes are reflected into left-moving ones and vice versa.
Since there is no scale in this problem, the boundary conditions
must be conformally-invariant; in the Ising model, this means
that the Ising spins can either have free or fixed boundary
conditions. When there is an even number of quasiparticles
in the bulk and their combined topological qubit has a fixed
fermion number parity, there are fixed boundary conditions
at both ends of the droplet. When there is an odd number
of quasiparticles in the bulk, there is a free boundary condition
at one end of the strip and a fixed boundary condition at
the other end. A Majorana fermion acquires a minus sign
when it goes around a $\sigma$; thus an odd number of $\sigma$
particles can change anti-periodic boundary conditions
to periodic. (For even $e/4$ quasiparticle numbers, every branch cut
can begin and end at a bulk quasiparticle, and no branch cuts need
cross the edge.) The branch cut emanating from a $\sigma$
can be moved anywhere we like by a $\mathbb{Z}_2$
gauge transformation. The most convenient place for
our purposes is one of the ends of the squashed droplet;
at this end, the Ising spin has free boundary condition.
By a $\mathbb{Z}_2$ gauge transformation, we could move the
branch cut to the other end. Interchanging
the free and fixed ends in this manner is simply a
Kramers-Wannier duality transformation. For details,
see Ref. \onlinecite{Fendley09}.

To apply this perspective to a two point-contact
interferometer, we will assume that ${x_0}=0$
and ${x_a}=-{x_b}$, which we can arrange by a conformal
transformation. Then, we fold the interferometer
about the point $x=0$, as depicted in Fig. \ref{fig:interfero}.
As a result, the Majorana fermion field on the bottom edge,
${\psi_1}(x)$, which was purely a right-moving field on the
line $-\infty<x<\infty$ now has both right- and left-moving
components, $\psi_{1R}(x)=\psi_{1}(x)$ and
$\psi_{1L}(x)=\psi_{1}(-x)$, on the half-line $x>0$.
The same holds for the top edge. For bulk-edge
coupling $h=0$, there is no scale in this problem, so
the boundary conditions at $x=0$ must be conformally-invariant.
If there is an odd-number of $e/4$ quasiparticles in the bulk,
then there will be a branch cut and, again, we are free to
put his branch cut wherever we like. As shown in
Fig. \ref{fig:interfero}, we will run the branch cut through
the bottom edge at the point $x=0$. Thus, the folded bottom
edge is a semi-infinite Ising model with free boundary condition
at $x=0$ while the folded top edge is
a semi-infinite Ising model with fixed boundary condition
at $x=0$.

In computing the interference term in the backscattered
current, we face expressions such as 
$$
\langle {\left[{\sigma_1}({x_a},0)\,{\sigma_1}({x_b},t)\right]_\psi}\rangle
=\langle {\left[\sigma_{1R}({x_a},0)\,\sigma_{1L}({x_a},t)\right]_\psi}
\rangle
$$
(recall that ${x_a}=-{x_b}$). According to Cardy's analysis
\cite{Cardy89}, this product of right- and left-moving
$\sigma$ fields can be combined into a single non-chiral
Ising spin field. For a given boundary condition, a non-chiral
one-point function can be expressed in terms of a chiral
two-point function in a definite fusion channel.
(In general, it is a linear combination over fusion channels,
but in the Ising case, it is a unique fusion channel.)
For free boundary condition, a non-chiral spin field
can be written as the product of chiral spin fields which
fuse to $\psi$:
\begin{equation}
\label{eqn:chiral-nonchiral-free}
\langle{\left[\sigma_{1R}({x_a},0)\,\sigma_{1L}({x_a},\tau)\right]_\psi}\rangle
= \langle{\sigma_1}(z,\overline{z})\rangle_{\rm free}
\end{equation}
while, for fixed boundary condition, a non-chiral spin field
can be written as the product of chiral spin fields which
fuse to $1$:
\begin{equation}
\label{eqn:chiral-nonchiral-fixed}
\langle{\left[\sigma_{2R}({x_a},0)\,\sigma_{2L}({x_a},\tau)\right]_1}\rangle
= \langle{\sigma_2}(z,\overline{z})\rangle_{\rm fixed}
\end{equation}
On the right-hand-sides of these equations,
the non-chiral spin fields are functions of
$z$, $\overline{z}$, which may be treated as formally
independent variables. For the computation of the
current, we take $z=i{x_a}$ and $\overline{z}={v_n}\tau-i{x_a}$.

Equations \ref{eqn:chiral-nonchiral-free} and
\ref{eqn:chiral-nonchiral-fixed} can be understood
intuitively following the discussion of Ref. \onlinecite{Fendley09}.
For odd quasiparticle number, there should be no
branch cut anywhere, so that $\psi_{1R}(0)=\psi_{1L}(0)$
and $\psi_{2R}(0)=\psi_{2L}(0)$. Meanwhile,
free and fixed boundary conditions correspond to
${\psi_R}(0)=\pm {\psi_L}(0)$. The slight subtlety is that
${\psi_R}(0)={\psi_L}(0)$ corresponds to free boundary condition
and ${\psi_R}(0)=-{\psi_L}(0)$ corresponds to fixed boundary
condition if the boundary of the Ising model is on the
upper-half-plane and the boundary is the real axis.
On the bottom edge, this is precisely the identification 
which leads to Eq. \ref{eqn:chiral-nonchiral-free}.
However, in conformally mapping the upper-half-plane
to a strip, an additional minus sign enters so that, on the
bottom edge, $\psi_{1R}(0)=\psi_{1L}(0)$ corresponds to 
fixed boundary condition as in Eq. \ref{eqn:chiral-nonchiral-fixed}.

The one-point function of the spin field is non-zero
for fixed boundary condition.
\begin{equation}
\langle{\sigma_2}(z,\overline{z})\rangle_{\rm fixed} =
\frac{1}{\left(z-\overline{z}\right)^{1/8}}
\end{equation}
However, for free boundary condition
\begin{equation}
\langle{\sigma_1}(z,\overline{z})\rangle_{\rm free} = 0
\end{equation}
Thus, when these two correlation functions are multiplied
together in the computation of the interference term,
we obtain a vanishing result, as expected for an odd number of
quasiparticles \cite{Bonderson06a,Stern06,Overbosch07,Bishara08}.

While the fixed boundary condition is stable, the free boundary
condition is unstable to perturbation by a boundary
magnetic field, which causes a flow to fixed boundary
condition. As discussed in Ref. \onlinecite{Fendley09}, the
boundary magnetic field perturbation \cite{Cardy89,Chatterjee94}
is precisely the coupling of a bulk zero mode to the edge
in Eq. \ref{eqn:bulk-edge-coupling}. Since the action remains
quadratic, even with this perturbation, it is possible to
solve it exactly to determine its effect.

As a result of the folding procedure,
(\ref{eqn:bulk-edge-coupling}) now becomes
\begin{equation}
S^{\rm folded}_\text{bulk-edge} = \int dt\,
\left({\psi_0}{\partial_t}{\psi_0} + 
ih\,{\psi_0}\,
[\psi_{1R}(0)+\psi_{1L}(0)]\right)
\end{equation}
The equations of motion
for $\psi_0$, $\psi_{1R}$ and $\psi_{1L}$ at
$x=0$ are \cite{Fendley09}
\begin{eqnarray}
\label{eqn:Majorana-eom}
2\partial_t \psi_0 &=& ih[\psi_{1R}(0) + \psi_{1L}(0)]\ , \\
iv_n \psi_{1R}(0) &=& iv_n \psi_{1L}(0) + h\psi_0\ , \\
iv_n \psi_{1L}(0) &=& iv_n \psi_{1R}(0) - h\psi_0 \ .
\end{eqnarray}
Consequently, the Fourier transforms satisfy:
\begin{equation}
\psi_R(x=0,\omega) =   \frac{\omega + i\omega_0} {
\omega - i \omega_0} \cdot \psi_L(x=0, \omega) ,
\label{bcres}
\end{equation}
Thus, we see that that a branch cut develops
at low energies, $\omega\ll {h^2}/2{v_n}$, so that
it is as if the $e/4$ quasiparticle is absorbed by the edge
(thereby switching the quasiparticle number
parity to even, which requires a branch cut)
at least as far as its non-Abelian topological properties
are concerned. According to the correspondence
of the previous paragraph, the emergence of a branch cut
is equivalent to the flow from free to fixed boundary condition.
In the presence of a boundary magnetic field perturbation of
the free boundary condition, we will denote the right-hand-side of
Eq. \ref{eqn:chiral-nonchiral-free} by
\begin{equation}
\label{eqn:chiral-nonchiral-h}
\langle{\left[\sigma_{1R}({x_a},0)\,\sigma_{1L}({x_a},t)\right]_\psi}
\rangle^{}_{{S_0}+S_{\rm bulk-edge}}
= \langle{\sigma_1}(z,\overline{z})\rangle_{\rm h}
\end{equation}

To summarize, according to the arguments of this
section, we can write
\begin{multline}
\label{eqn:P-Ising-mapping}
\tilde{P}({\omega^{}_J}) = \int_{-\infty}^\infty \!\!
dt\, e^{i{\omega^{}_J} t}\,
\left[(2{x_a})^2-({v_c}t)^2+\delta\,\text{sgn}(t)\right]^{-1/8}
\:\times\\
\langle{\sigma_1}(z,\overline{z})\rangle_{\rm h}\,\,
\langle{\sigma_2}(z,\overline{z})\rangle_{\rm fixed},
\end{multline}
with $z=i{x_a}$ and $\overline{z}={v_n}\tau-i{x_a}$
analytically-continued to real time $t$.
The final factor in the first line comes from the
bosonic charged mode correlation functions.
From (\ref{eqn:P-Ising-mapping}),
the interference term in the backscattered
current is obtained via Eq. \ref{eqn:interference-term}.

Although the action ${S_0}+S_{\rm bulk-edge}$ is quadratic,
the desired correlation function,
$\langle{\sigma_1}(z,\overline{z})\rangle_{\rm h}$,
is complicated because the spin field does not have
a simple relationship to the Majorana fermion -- since
it creates a branch cut for the Majorana
fermion, it is non-local with respect to it. Nevertheless, it
can be computed exactly, as shown by Chatterjee and Zamolodchikov
\cite{Chatterjee94}. We recapitulate their method in Appendix
\ref{sec:Chatterjee-Zamo}.

In order to compute the current, we need to combine (\ref{eqn:P-Ising-mapping})
with the result for $\langle{\sigma_1}(z,\overline{z})\rangle_{\rm h}$
described in Appendix \ref{sec:Chatterjee-Zamo}.
However, there is one small subtlety:
the correlation function (\ref{eqn:sigma-crossover2})
is an imaginary-time expression which needs to be analytically
continued to real time. This is a little delicate because the
imaginary-time correlation function of chiral $\sigma$ fields
$\sigma_{1R}({x_a},0)\,\sigma_{1L}({x_a},\tau) \sim {\sigma_1}(z,\overline{z})$
is multi-valued.
The more serious multi-valuedness, associated with
non-Abelian statistics, occurs in higher-point functions,
and is handled by fixing the fusion channel \cite{Fendley06,Fendley07a}.
However, even for fixed fusion channel,
there is a phase ambiguity. This type of ambiguity is characteristic
of chiral order and disorder operators, including exponentials
of chiral bosons $e^{i\phi_R}$. In the classical statistical mechanics
context, this ambiguity disappears since the combination
${\sigma_1}(z,\overline{z})$ is single-valued when $z=(\overline{z})^*$
(and the non-Abelian ambiguity, when it is present, is eliminated by
taking a single-valued sum over fusion channels). In the
quantum context, the particular combinations of such correlation functions
which enter physical quantities are single-valued.
A simple example is the case of fixed boundary condition,
which is the $h\rightarrow\infty$ limit of (\ref{eqn:sigma-crossover2}).
Then ${\sigma_1}(i{x_a},{v_n}\tau-i{x_a})=({v_n}\tau-2i{x_a})^{1/8}$.
This is multivalued. However, in
the computation of the current, ${\sigma_2}(z,\overline{z})$ only
enters in the combination $\langle{\sigma_1}(z,\overline{z})\rangle_{\rm fixed}\,\,
\langle{\sigma_2}(z,\overline{z})\rangle_{\rm fixed}=(({v_n}\tau)^2+(2{x_a})^2)^{1/8}$.
The real-time correlation function will have real, physical
singularities on the light cone $2{x_a}=\pm{v_n}t$, but it will not
be multi-valued.
Thus, we can avoid ambiguities by forming the single-valued combinations
which enter into physical quantities before continuing to real time.
Fortunately, the typical route to calculating a response
function, namely to compute the correlation function in imaginary time
and then make the substitution
$i\omega\rightarrow {\omega_J}+i\delta$ (e.g. in Kubo formula
calculations using Matsubara frequencies),
deals only with such combinations.

\section{Crossover Scaling Function for the
Current through the Interferometer}

As explained in Appendix \ref{sec:Chatterjee-Zamo},
the Ising spin field one-point function
for finite-boundary magnetic field takes the
form:
\begin{equation}
\label{eqn:sigma-crossover2}
\langle {\sigma_1}(w,\overline{w})\rangle = 
\lambda^{1/2}\,2^{1/4}\,y^{3/8}\,\,
U\!\left(\mbox{$\frac{1}{2}$},1,y\right)
\end{equation}
where $y=-i\lambda(w-\overline{w})=\lambda(2{x_a}+i{v_n}\tau)$,
$\lambda={h^2}/2{v_n^2}$
and $U\!\left(\mbox{$\frac{1}{2}$},1,y\right)$
is the confluent hypergeometric function of the second
kind, discussed briefly in Appendix \ref{sec:Chatterjee-Zamo}.
As discussed in the previous section, we combine
(\ref{eqn:P-Ising-mapping}) and
(\ref{eqn:sigma-crossover2}), remaining in
imaginary time. We have 
\begin{multline}
\label{eqn:P-Ising-mapping-2}
\tilde{P}(\omega) = \int_{-\infty}^\infty \!\!
dt\, e^{i\omega \tau}\,
\left[({v_c}\tau)^2 + (2{x_a})^2\right]^{-1/8}
\:\times\\
\left[{v_n}\tau+2i{x_a}\right]^{-1/8}\,
\left[{v_n}\tau-2i{x_a}\right]^{3/8}\:\times\\
\lambda^{1/2}\,2^{1/4}\,\cdot\,
U\!\left(\mbox{$\frac{1}{2}$},1,\lambda(2{x_a}+i{v_n}\tau)\right)
\end{multline}
As mentioned in the previous section, individual factors
in the integral have ambiguities. However, their combination
does not. Thus, we use (\ref{eqn:P-Ising-mapping-2}) to compute 
$\tilde{P}(\omega)$ and then take $i\omega\rightarrow {\omega_J}+i\delta$.

If we consider low voltages, ${v_n}\gg 2{x_a}\omega_J$,
then we can drop the ${x_a}$ dependence, so
that Eq. \ref{eqn:P-Ising-mapping-2} simplifies considerably.
Using the integral representation of $U(\mbox{$\frac{1}{2}$},1,y)$,
valid for $\text{Re}(y)>0$, given in Eq. \ref{eqn:hyper-integral-rep},
$$
U(a,b,y)=\frac{1}{\Gamma(a)}{\int_0^\infty}\!dt\,
e^{-ty}\, t^{a-1}\, (1+t)^{b-a-1}
$$
the integral can be performed:
\begin{align}
\label{eqn:P-simple-result}
\tilde{P}({\omega}) &=\nonumber\\
& \int_{-\infty}^\infty \!\!
dt\, e^{i\omega \tau}\,\bigl(\mbox{$\frac{v_n}{v_c}$}\bigr)^{1/4}
\lambda^{1/2}\,2^{1/4}\,
U\!\left(\mbox{$\frac{1}{2}$},1,\lambda(2{x_a}+i{v_n}\tau)\right)
\nonumber\\
&= 2\pi^{1/2}\,\lambda^{1/2}\,\bigl(\mbox{$\frac{v_n}{v_c}$}\bigr)^{1/4}\,
\frac{\theta(\omega)}{\left[\omega
({v_n}\lambda+\omega)\right]^{1/2}}
\end{align}
Substituting $i\omega\rightarrow {e^*}V+i\delta$, we see that
for ${e^*}V\gg{v_n}\lambda$, the interference term
has voltage dependence $\sim h/V$, as calculated
perturbatively in Ref. \onlinecite{Overbosch07}.
However, for ${e^*}V\ll{v_n}\lambda$,
the interference term is $\sim 1/V^{1/2}$ -- i.e. scales the same
way with voltage as the separate contributions from each
point contact, $I_a$ and $I_b$ -- and is independent
of $h$. Thus, as expected, the Ising model crosses over from
free to fixed boundary conditions, which is reflected in
the interferometer as a crossover from odd to even quasiparticle
numbers. The total current, including the individual contact
and interference terms is:
\begin{multline}
\label{eqn:total-current-cross}
I_{\rm total} =  \pi^{1/2} {e^*} 
\left(|\Gamma_a^2|+|\Gamma_b|^2\right)({v_n}{v_c})^{-1/4}\,
\text{sgn}(V) \left({e^*}|V|\right)^{-\frac{1}{2}}\\
+ 
2\pi^{1/2}{e^*} \,
\bigl(\mbox{$\frac{v_n}{v_c}$}\bigr)^{1/4}\lambda^{1/2}\,\,
\text{Re}\!\left\{\frac{{\Gamma^{}_a}{\tilde{\Gamma}_b^*}\:\:
\text{sgn}(V)}{\left[{e^*}V
(i{v_n}\lambda+{e^*}V)\right]^{1/2}}\right\}
\end{multline}
This regime, ${v_n}\gg 2{x_a}\omega_J$,
is accessible to experiments (e.g. those of
Refs. \onlinecite{Willett09,Zhang09}) 
since $2{x_a}\approx 1 \mu\mbox{m}$ compared to
${v_n}/{e^*}V \sim 10\mu\mbox{m}$  for $V\approx 1\mu\mbox{V}$.
In this regime, ${v_n}/{x_a}$ is much larger than the other energy scales
and is unimportant for the crossover between odd and even quasiparticle
numbers, which occurs when ${v_n}\lambda={h^2}/2{v_n}$ is increased
until it approaches ${e^*} V$, as may be seen from (\ref{eqn:total-current-cross}).
(Or, conversely, when the voltage is decreased
until it approaches ${h^2}/2{v_n}$).

\begin{figure}[t!]
\includegraphics[width=3in]{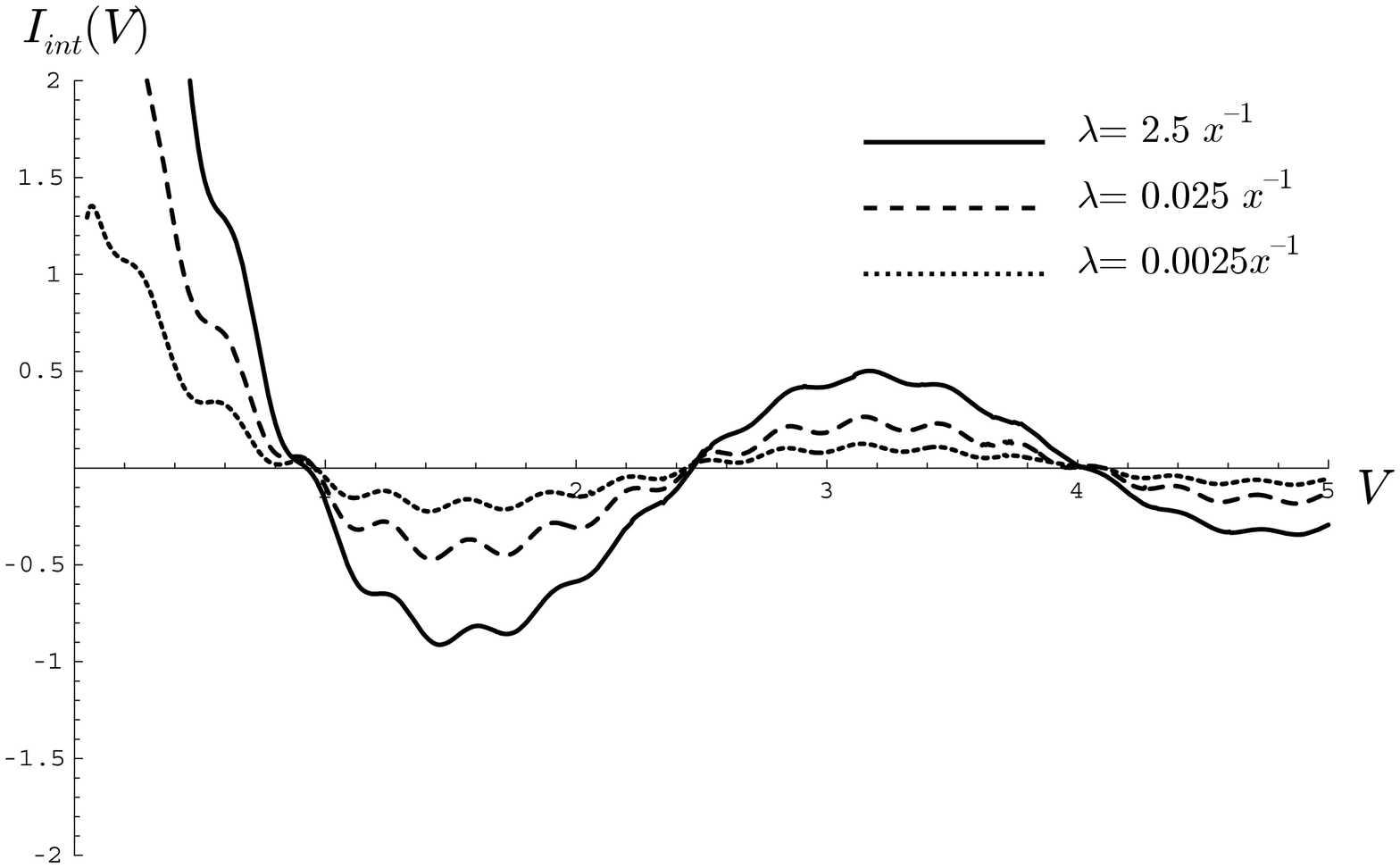}
  \includegraphics[width=3in]{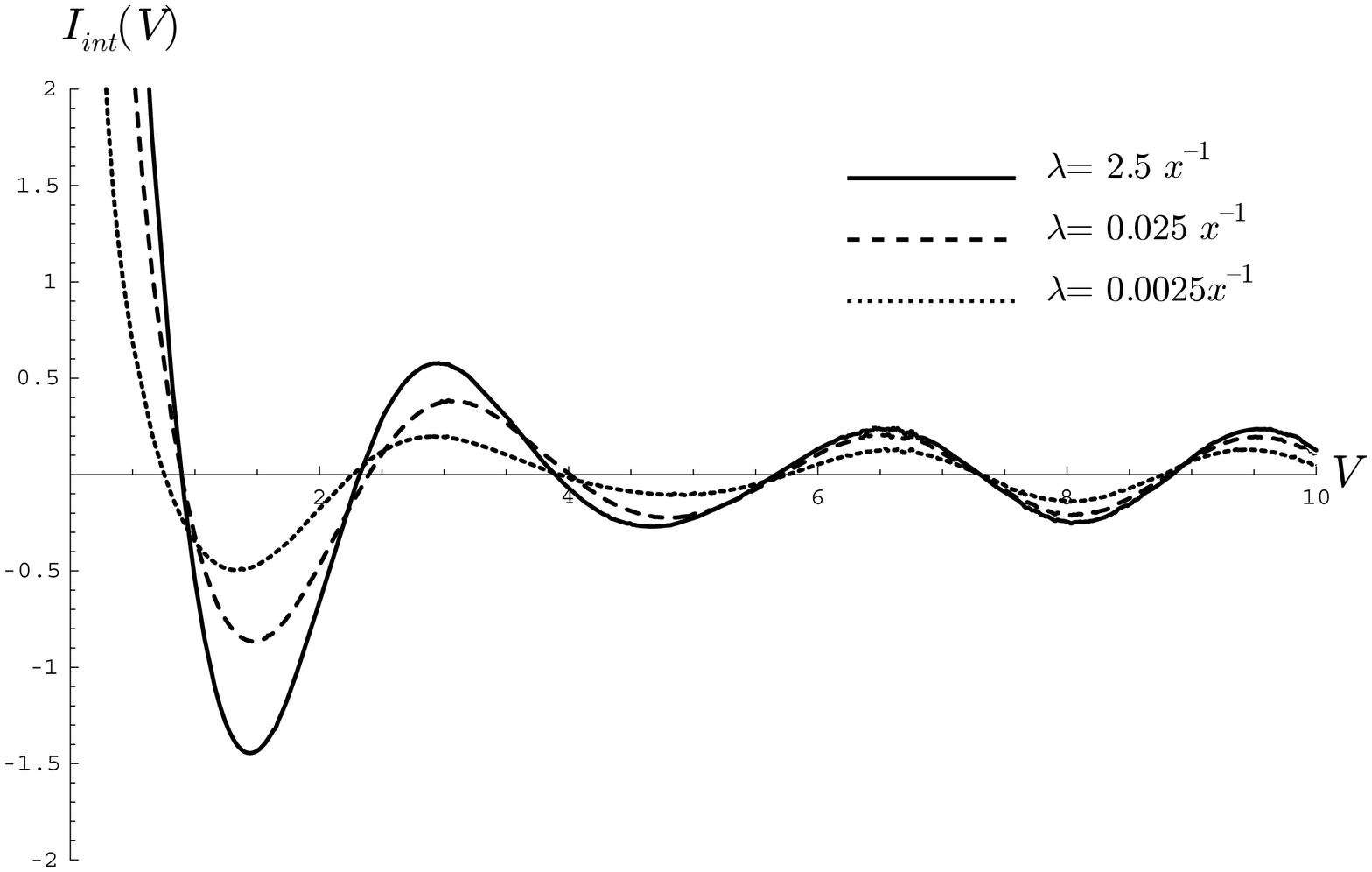}
  \caption{The interference term in the current as a function of applied
  voltage at low temperature, for $v_n= 0.1\, v_c$ (upper panel) and
  $v_n= 0.75\, v_c$ (lower panel).
  The three curves in each are three different bulk-edge coupling strengths
  $\lambda={h^2}/2{v_n^2}$, given in units of $1/{x_a}$, the inverse of the
  separation between the point contacts.}
  \label{fig:finite-voltage}
\end{figure}

\begin{figure}
  \includegraphics[width=3in]{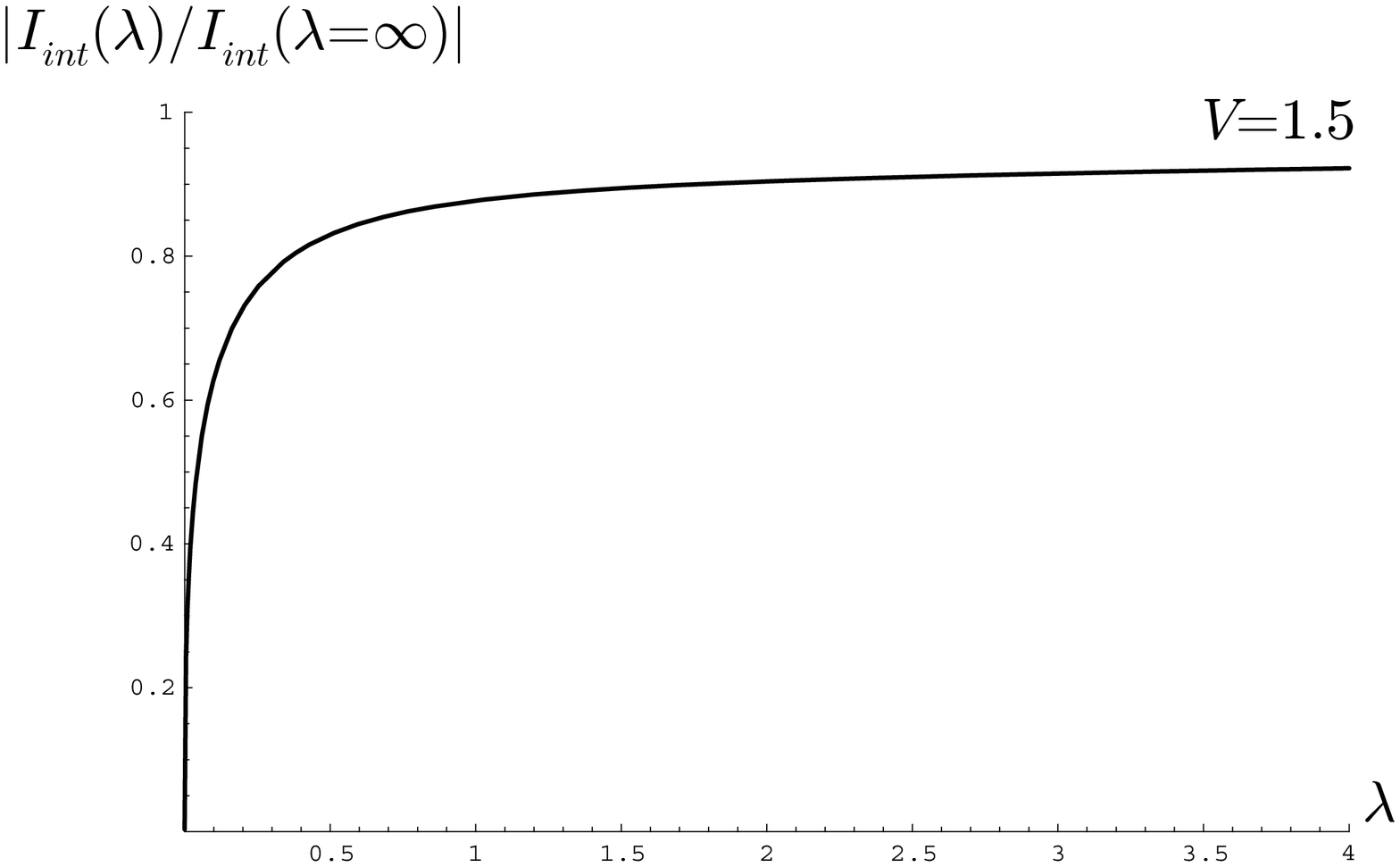}
   \includegraphics[width=3in]{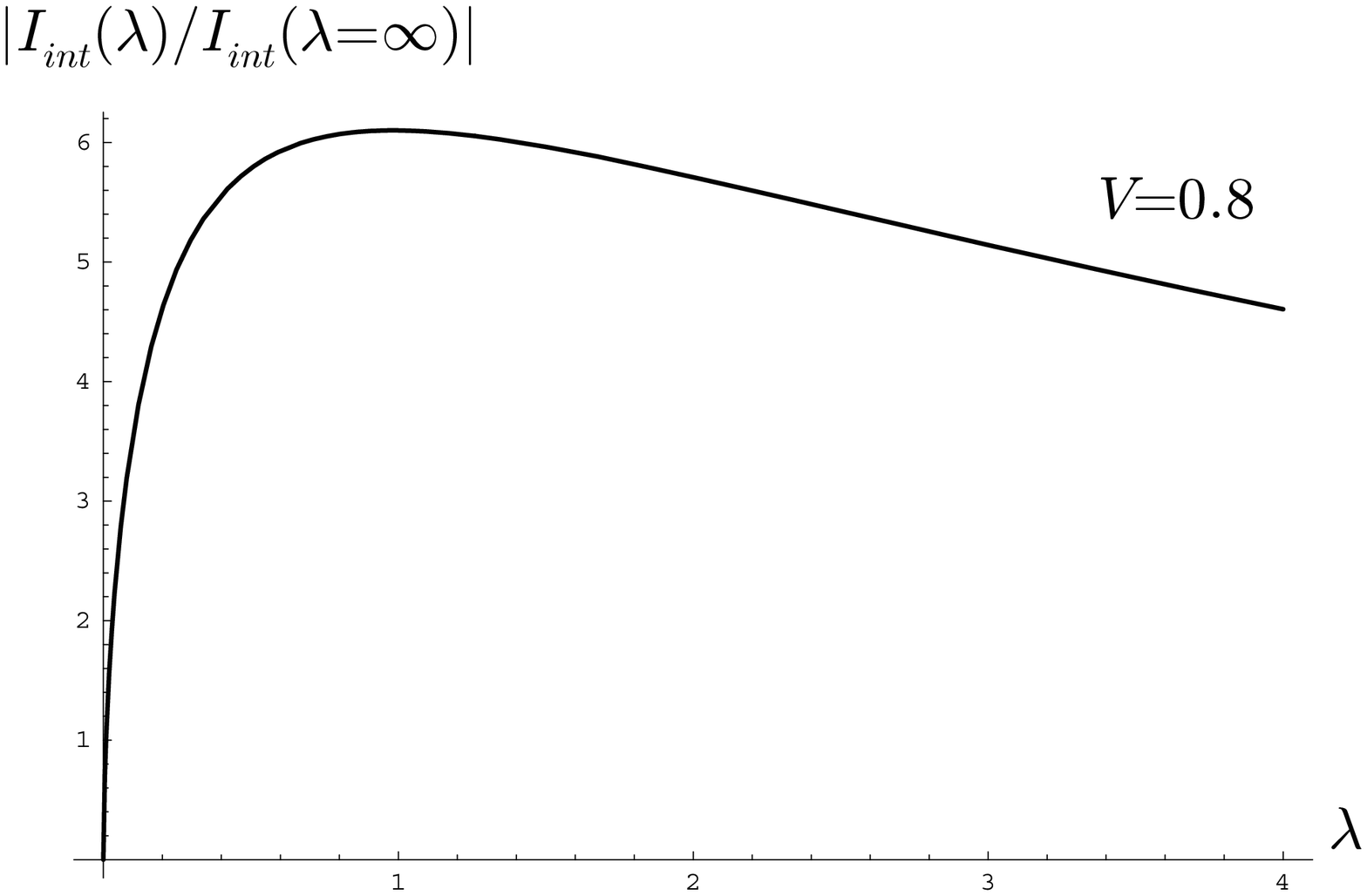}
  \caption{The amplitude of the interference term for fixed applied voltage
  ${e^*}V>{v_n}/2{x_a}$ as a function of bulk-edge coupling $\lambda$.
  $\lambda$ and ${e^*}V$ are measured in units of ${v_n}/2{x_a}$.
  The upper panel is away from a nodal point; the amplitude asymptotes
  its large $\lambda$ value at $\lambda\approx{v_n}/2{x_a}$. This
  represents the behavior of the envelope of the interference term;
  it agrees with the numerical calculation of Ref. \onlinecite{Rosenow07b}.
  The lower panel is for $V$ near a nodal point in $I_{\rm int}$; the amplitude
  varies non-monotonically with $\lambda$ because the nodes move as
  $\lambda$ is varied.}
  \label{fig:amp-lambda}
\end{figure}

However, for larger voltages $V\approx 10-100\mu$V and/or
larger interferometers $2{x_a}\approx 10 \mu\mbox{m}$, which are also experimentally accessible (see Ref. \onlinecite{Zhang09}),
oscillations with voltage will be observed for
even quasiparticle number, as shown, for instance,
in Fig. 3 of Ref. \onlinecite{Bishara08}. There are `fast' oscillations
with period in $V$ given roughly by
$\frac{16\pi}{e|x_1-x_2|}(1/v_n+1/v_c)^{-1}$ and `slow'
ones with larger period, $\frac{16\pi}{e|x_1-x_2|}(1/v_n-1/v_c)^{-1}$.
For an odd number of quasiparticles in the interferometer,
one of which is close to an edge, oscillations are seen,
but they are small for $\lambda\ll 1/{x_a}$. For $\lambda\gg x_a$,
on the other hand, the interference term in the
current approaches the even quasiparticle number case,
as shown in Fig. \ref{fig:finite-voltage}. However, if ${e^*}V$
is not much smaller than ${v_n}/{x_a}$, this will occur in a more
complicated way than in Eq. \ref{eqn:total-current-cross}. For example,
the nodes in the oscillations move as $\lambda$ is varied. Thus, if the voltage
is near a nodal point in $I_{\rm int}(V)$, the current
will not approach its $\lambda\rightarrow\infty$
value monotonically, as shown in Fig. \ref{fig:amp-lambda}.

Since the preceding formulas were computed perturbatively
in the inter-edge backscattering operators, they are only
valid for voltages which are not too small, i.e. so
long as $|\Gamma_{a,b}|^2 |{e^*}V|^{-1/2}\ll 1$. Thus,
the crossover described above will be observable
if there is a regime $|\Gamma_{a,b}|^4 \ll {e^*}V \ll \frac{h^2}{v_n}$
(here, we have substituted $\lambda={h^2}/{2v_n^2}$).
However, it is possible to go to voltages lower than $|\Gamma_{a,b}|^4$,
while still remaining in the weak-backscattering regime,
if the temperature is finite, since ${k_B}T$ will then
cut off the flow of $\Gamma_{a,b}$.

Finite-temperature correlation functions can be obtained from
zero-temperature ones such as (\ref{eqn:sigma-crossover2}) by
a conformal map from the half-plane to the half-cylinder. This
amounts to the following substitution.
\begin{equation}
\label{TempSub}
\delta+i(t\pm x/v)\rightarrow
\left({\sin\left(\pi T(\delta+i(t\pm x/v))\right)}\right){\pi T}
\end{equation}
Since the charged and neutral mode velocities are
different, we apply such a substitution separately to
the charged and neutral sectors of the theory, which
we can do only because they are decoupled in the
weak-backscattering limit.
The $I-V$ curves shown in Fig. \ref{fig:finite-voltage}
are computed at small but non-zero temperature.
(Since the temperature acts as an
infrared regulator, it calculationally convenient.)

\section{Discussion}

Until very recently, the evidence that the $\nu=5/2$
state is in the universality class of either the Moore-Read
Pfaffian state \cite{Moore91} or the anti-Pfaffian state
\cite{LeeSS07,Levin07} was derived entirely from
numerical solutions of small systems \cite{Morf98,Rezayi00}.
However, recent point-contact tunneling \cite{Radu08}
and shot-noise \cite{Dolev08} experiments
are consistent with these non-Abelian states. 
Even more recently, measurements \cite{Willett09}
with a two point-contact interferometer appear
consistent with the odd-even effect \cite{Bonderson06a,Stern06}.
In this paper, we have computed how the coupling of
a bulk $e/4$ quasiparticle to the edge leads to a crossover
between the odd and even quasiparticle number regimes.
Our results may be relevant to the transition regime
between different quasiparticle numbers in the experiment
of Ref. \onlinecite{Willett09}. We have made specific
predictions in Eqs. \ref{eqn:total-current-cross} for how the interference term
scales with voltage and temperature when there is appreciable
bulk-edge coupling. If the transition regions can be studied
as a function of temperature and voltage, a comparison may be possible.

Following Willett et al. \cite{Willett09},
let's assume that $\Delta\Phi= c\Delta V_s$ for some constant $c$,
where $V_s$ is the sidegate voltage.
When a bulk quasiparticle is close to the edge, $h$ will be large,
the dependence of $h$ on $V_g$ will be complicated. However,
for some range of $V_s$, $h$ will be
approximately $h\sim e^{-r/\xi}$, where
$r$ is the distance to the edge and $\xi$ is a length
scale corresponding to the size of the Majorana bound
state at the bulk $e/4$ quasiparticle. If $r={r_0}-b{V_s}$,
then one might expect $h={h_0} e^{b{V_s}}$ for some $h_0$, $b$.
We take Willett {\it et al.}'s $c$ and choose ${h_0},b$
so that the quasiparticle is effectively absorbed by the edge
after $\sim 5$ periods. To compare with the results
of Ref. \onlinecite{Willett09},
we must add to (\ref{eqn:total-current-cross})
the contribution of charge $e/2$
quasiparticles:
\begin{multline}
\label{eqn:half-e-contribution}
I^{\left(e/2\right)} = \frac{e}{2}\, \frac{2 \pi}{v_{c}}\Bigl(| \Gamma_a^{e/2}|^{2}
+ \left| \Gamma_b^{e/2} \right|^{2}\Bigr)\\
+ 2\cdot\frac{e}{2}\,\frac{2 \pi}{v_{c}}\,\,
\text{Re}\left\{ \Gamma_a^{e/2} (\Gamma_b^{e/2})^* 
e^{2\pi i(\frac{\Phi}{2\Phi_0}-\frac{{n_q}}{4})}
\right\}
\end{multline}
\begin{figure}
  \includegraphics[width=3in]{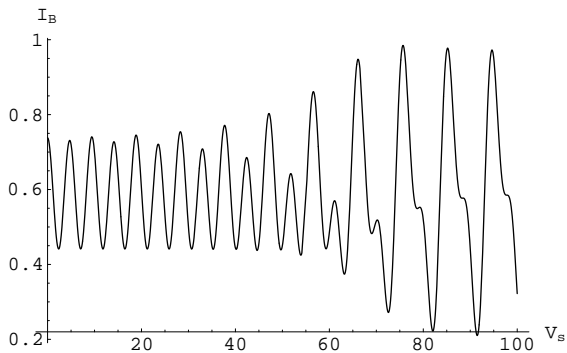}
   \caption{The backscattered current $I_B$ (normalized
   by its maximum value) as a function of
   sidegate voltage $V_s$ (in mV), assuming that
   there is an odd number of $e/4$ quasiparticles in
   the interference loop, one of which is close to the edge.
   The bulk-edge coupling is assumed to vary with sidegate
   voltage as $h={h_0} e^{b{V_s}}$ while the flux through the loop
   varies as $\Delta\Phi= c\Delta V_s$. A contribution to the current from
   charge $e/2$ quasiparticles (with period $2\Phi_0$) is also included
   with an amplitude which is approximately half the large-$h$
   limit of the amplitude of the period $4\Phi_0$ oscillations.}
  \label{fig:crossover-fit}
\end{figure}
If we assume that the backscattered
$e/2$ quasiparticle contribution to the current is
half as large as the (large-$\lambda$ limit of the)
$e/4$ contribution (although this is a somewhat
questionable assumption in general, it may hold
over a range of temperatures, see e.g. Ref. \onlinecite{Bishara09}),
then we can obtain the total backscattered current
as a function of $V_s$, as shown in Fig. \ref{fig:crossover-fit}.
A striking feature of this plot is that, for large $h$,
the $e/2$ oscillation (with period $2\Phi_0$) is masked by the larger
$e/4$ oscillation (with period $4\Phi_0$).
We emphasize that {\it the amplitude of the
$e/2$ oscillation is not changing with $V_s$},
as may be seen from Eq. \ref{eqn:half-e-contribution}; the apparent
suppression of the $e/2$ oscillation as the amplitude of
the $e/4$ oscillation increases is illusory.
This is reminiscent of a salient feature of
Willett {et al.}'s \cite{Willett09} data: in the regions in which
$e/4$ oscillations are visible, $e/2$ oscillations are often
barely, if at all, visible. The apparent disappearance of the $e/2$
oscillation in Fig. \ref{fig:crossover-fit} results, in part, from the $\pi/4$
phase shift of the $e/4$ oscillation in Eq. \ref{eqn:total-current-cross},
which helps it submerge the smaller $e/2$ oscillation.
This suggests that some of the regions in Willett {et al.}'s \cite{Willett09}
data which have been interpreted as having an even number of quasiparticles
in the interference loop because of the presence of
$e/4$ oscillations may, in fact, have an odd number of quasiparticles
in the loop, one of which is close enough to the edge that $\lambda$
is large and the quasiparticle is effective absorbed, as in
Fig. \ref{fig:crossover-fit} for ${V_s}>70\,\text{mV}$. If this is true,
then we would expect that, if ${v_n}\lambda$ is much smaller than
the energy gap $\Delta$ of the $\nu=5/2$ quantum Hall state,
then as the temperature is raised above ${v_n}\lambda$, the $e/4$
oscillation will disappear (due to both the effective decoupling of
one of the quasiparticles from the edge and also thermal smearing
of the $e/4$ oscillation \cite{Wan08,Bishara09}) and
only the $e/2$ oscillation will be left.
Finally, it is amusing to note that if we assume that
$h$ increases even more sharply with ${V_s}$,
e.g. $h={h_0} e^{b{V_s^2}}$, then the $e/2$ oscillation
seemingly disappears even more suddenly.

In this paper, we have focussed on the case of an
odd number of quasiparticles in the bulk, one of which
is coupled to the edge. The case of an even number
of quasiparticles is qualitatively different: in the absence
of bulk-edge coupling, interference is observed, with a 
phase which is determined by the combined topological
state of the quasiparticles in the bulk. For instance,
when there are two quasiparticles
in an interference loop, they form a qubit (or half a qubit,
if four quasiparticles with total topological charge $1$
are used to represent a qubit) \cite{DasSarma05,Nayak08}.
Bulk-edge coupling then leads to errors in this qubit and,
over long enough time scales, to the disappearance of
this qubit as one of the bulk quasiparticles is absorbed
by the edge. This non-equilibrium problem will
be the subject of a separate work.

\begin{figure}
  \includegraphics[width=3in]{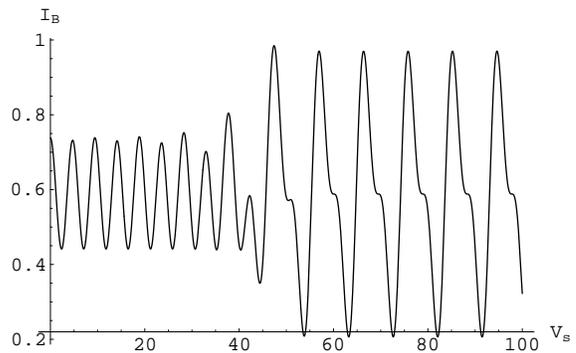}
   \caption{If we assume that $h$ increases more sharply
   with $V_s$, e.g. $h={h_0} e^{bV_s^2}$, then the apparent
   disappearance of $e/2$ oscillations (with period $2\Phi_0$)
   as $h$ increases with $V_s$ is even more dramatic.}
  \label{fig:crossover-fit2}
\end{figure}

\appendix

\section{Method of Chatterjee and Zamolodchikov}
\label{sec:Chatterjee-Zamo}

As a consequence of the mapping introduced in
Section \ref{sec:mapping}, we can reduce the problem of finding the
interference term in the current backscattered in the
interferometer to that of finding
$\langle{\sigma_1}(z,\overline{z})\rangle_{\rm h}$ with
$z,\overline{z}$ independent variables. 
Following Chatterjee and Zamolodchikov \cite{Chatterjee94},
we note that Eqs. \ref{eqn:Majorana-eom} imply that
\begin{equation}
\label{eqn:continuation}
\left({\partial_t}+\frac{ih^2}{2v_n}\right)\!\psi_{1R}(0,t)
= \left({\partial_t}-\frac{ih^2}{2v_n}\right)\!\psi_{1L}(0,t)
\end{equation}
Thus, the combination on the right-hand-side of
(\ref{eqn:continuation}) is the continuation of the left-hand-side,
simply reflected back by $x=0$, analogous to ${\psi_R}(0)={\psi_L}(0)$
for free boundary condition. Chatterjee and Zamolodchikov
\cite{Chatterjee94} observe that this fact, which can
be written in complex notation as
\begin{equation}
\left[({\partial_z}+i\lambda)\psi_{1R}-
(\partial_{\overline{z}}-i\lambda)\psi_{1L}\right]_{z=\overline{z}} = 0
\end{equation}
where $\lambda={h^2}/2{v_n^2}$, allows them to treat
$({\partial_z}+i\lambda)\psi_{1R}$ as a free field
unaffected by the boundary interaction.

Now, consider the quantity
$\langle\psi_{1R}(z) {\mu_1}(w,\overline{w}) \rangle$,
where ${\mu_1}(w,\overline{w})$ is the disorder operator
dual to ${\sigma_1}(w,\overline{w})$,
\begin{equation}
{\psi_R}(z)\cdot{\sigma_1}(w,\overline{w}) = \frac{\omega}{\sqrt{2}}\,
{\mu_1}(w,\overline{w}) + \ldots
\end{equation}
For fixed boundary condition, $\langle\psi_{1R}(z) {\mu_1}(w,\overline{w}) \rangle$ can be deduced by scaling and
the requirement of square root branch points
at $w$ and $\overline{w}$:
\begin{equation}
\left\langle\psi_{1R}(z)\, {\mu_1}(w,\overline{w}) \right\rangle_{\rm fixed}
= \frac{(w-\overline{w})^{3/8}}{(z-w)^{1/2}(z-\overline{w})^{1/2}}
\end{equation}
Thus, for fixed boundary condition,
\begin{multline}
\label{eqn:corr-form}
\left\langle({\partial_z}+i\alpha)\psi_{1R}(z)\,
{\mu_1}(w,\overline{w}) \right\rangle_{\rm fixed}
 (z-w)^{1/2}(z-\overline{w})^{1/2}
=\\
\frac{A(w,\overline{w})}{z-w} +
\frac{\overline{A}(w,\overline{w})}{z-\overline{w}}
+ B(w,\overline{w})
\end{multline}
holds for arbitrary $\alpha$ with
\begin{equation}
\label{eqn:A-barA-B}
A(w,\overline{w})=\overline{A}(w,\overline{w})=
iB(w,\overline{w})/2\alpha=
-\frac{1}{2}(w-\overline{w})^{3/8}
\end{equation}
For free boundary condition,
$\langle\psi_{1R}(z) {\mu_1}(w,\overline{w}) \rangle$
vanishes by fermion number parity, so (\ref{eqn:corr-form})
holds trivially with $A(w,\overline{w})=\overline{A}(w,\overline{w})=
B(w,\overline{w})=0$. 
For non-zero $h$, Eq. \ref{eqn:corr-form} must hold
for the special value $\alpha=\lambda$ since
$({\partial_z}+i\lambda)\psi_{1R}(z)$ is a free field, i.e.
this correlation function's only singularities are square root branch
points at $w$ and $\overline{w}$:
\begin{multline}
\label{eqn:corr-form-h}
\left\langle\,{\chi^{}_R}(z)
{\mu_1}(w,\overline{w}) \right\rangle_{h}
 (z-w)^{1/2}(z-\overline{w})^{1/2}
=\\
\frac{A(w,\overline{w})}{z-w} +
\frac{\overline{A}(w,\overline{w})}{z-\overline{w}}
+ B(w,\overline{w})
\end{multline}
where
\begin{equation}
{\chi^{}_R}(z)\equiv({\partial_z}+i\lambda)\psi_{1R}(z)
\end{equation}
but $A(w,\overline{w})$, $\overline{A}(w,\overline{w})$.
However, $B(w,\overline{w})$ in (\ref{eqn:corr-form-h})
will no longer have their free field forms (\ref{eqn:A-barA-B}).
In the $z\rightarrow w$ limit, (\ref{eqn:corr-form-h}) becomes
\begin{multline}
\label{eqn:corr-form-limit}
\left\langle\,{\chi^{}_R}(z)
{\mu_1}(w,\overline{w}) \right\rangle_{h} =
 (z-w)^{-3/2}\frac{A(w,\overline{w})}{(w-\overline{w})^{1/2}}\\\
- \frac{1}{2}(z-w)^{-1/2}\frac{A(w,\overline{w})}{(w-\overline{w})^{3/2}}
+ \frac{3}{8} (z-w)^{1/2}\frac{A(w,\overline{w})}{(w-\overline{w})^{5/2}}\\
 +(z-w)^{-1/2}\frac{\overline{A}(w,\overline{w})}{(w-\overline{w})^{3/2}}
 - \frac{3}{2}(z-w)^{1/2}\frac{\overline{A}(w,\overline{w})}
 {(w-\overline{w})^{5/2}}\\
+ (z-w)^{-1/2}\frac{B(w,\overline{w})}{(w-\overline{w})^{1/2}}
- \frac{1}{2}(z-w)^{1/2}\frac{B(w,\overline{w})}{(w-\overline{w})^{3/2}}
\end{multline}

On the other hand, the operator product expansion of
${\chi^{}_R}(z)\equiv({\partial_z}~+~i\lambda)\psi_{1R}(z)$ with
${\mu_1}(w,\overline{w})$ is determined by the
short-distance properties of the theory, i.e.
${\mu_1}$ is still simply the operator which creates a
branch cut for $\psi_{1R}$, even in the presence of a
boundary magnetic field. Thus, this OPE can be computed using
$S_0$:
\begin{multline}
\label{eqn:chi-sigma-OPE}
{\chi^{}_R}(z)\,\cdot\,
{\mu_1}(w,\overline{w}) =
 \frac{\overline{\omega}}{\sqrt{2}}
\Bigl[(z-w)^{-3/2}\left(-\mbox{$\frac{1}{2}$}{\sigma_1}(w,\overline{w})\right)\\
+ (z-w)^{-1/2}(2{\partial_w}+i\lambda){\sigma_1}(w,\overline{w})\\
+ (z-w)^{1/2}(4{\partial^2_w}+4i\lambda{\partial_w}){\sigma_1}(w,\overline{w})
+ \ldots\Bigr]
\end{multline}
where $\omega=e^{i\pi/4}$.

Taking the ground state expectation values of both sides of
(\ref{eqn:chi-sigma-OPE}) and comparing corresponding powers
of $z-w$ with (\ref{eqn:corr-form-limit}) leads to the equations:
\begin{equation}
\label{eqn:OPE-corr-a}
\frac{A(w,\overline{w})}{(w-\overline{w})^{1/2}} =
 \frac{\overline{\omega}}{\sqrt{2}}
\left[-\frac{1}{2}\left\langle{\sigma_1}(w,\overline{w})\right\rangle\right]
\end{equation}
\begin{multline}
\label{eqn:OPE-corr-b}
- \frac{1}{2}\frac{A(w,\overline{w})}{(w-\overline{w})^{3/2}}
 +\frac{\overline{A}(w,\overline{w})}{(w-\overline{w})^{3/2}} \:+\\
 \frac{B(w,\overline{w})}{(w-\overline{w})^{1/2}}
=  \frac{\overline{\omega}}{\sqrt{2}}
\left[(2{\partial_w}+i\lambda)\left\langle{\sigma_1}(w,\overline{w})\right\rangle\right]
\end{multline}
\begin{multline}
\label{eqn:OPE-corr-c}
\frac{3}{8}\frac{A(w,\overline{w})}{(w-\overline{w})^{5/2}}
 -\frac{3}{2}\frac{\overline{A}(w,\overline{w})}{(w-\overline{w})^{5/2}}\:-\\
\frac{1}{2} \frac{B(w,\overline{w})}{(w-\overline{w})^{3/2}}
=  \frac{\overline{\omega}}{\sqrt{2}}
\left[(4{\partial^2_w}+4i\lambda{\partial_w})\left\langle{\sigma_1}(w,\overline{w})\right\rangle\right]
\end{multline}
Substituting (\ref{eqn:OPE-corr-a}) into (\ref{eqn:OPE-corr-b}),
leads to
\begin{multline}
\label{eqn:OPE-corr-b'}
 B(w,\overline{w})+\frac{\overline{A}(w,\overline{w})}{(w-\overline{w})}
=\\  \frac{\overline{\omega}}{\sqrt{2}}(w-\overline{w})^{1/2}\!
\left(2{\partial_w} +
i\lambda-\frac{1}{4}\frac{1}{w-\overline{w}}\right)
\left\langle{\sigma_1}(w,\overline{w})\right\rangle
\end{multline}
and substituting (\ref{eqn:OPE-corr-b'}) into (\ref{eqn:OPE-corr-c})
leads to
\begin{multline}
\label{eqn:OPE-corr-c'}
-\frac{\overline{A}(w,\overline{w})}{(w-\overline{w})^{5/2}}
=  \frac{\overline{\omega}}{\sqrt{2}}(w-\overline{w})^{1/2}
\biggl[4{\partial^2_w} + 4i\lambda{\partial_w} +
\frac{1}{w-\overline{w}}{\partial_w}\\ +
\frac{1}{2}i\lambda\frac{1}{w-\overline{w}}
-\frac{1}{16}\frac{1}{(w-\overline{w})^2}\biggr]\!\!
\left\langle{\sigma_1}(w,\overline{w})\right\rangle
\end{multline}

Proceeding in a precisely analogous manner,
similar equations can be derived for the OPE and
correlation function of ${\chi^{}_L}(\overline{z})
\equiv(\partial_{\overline{z}}~-~i\lambda)\psi_{1L}(\overline{z})$
and ${\mu_1}(w,\overline{w})$, from which it
follows that:
\begin{equation}
\label{eqn:anti-OPE-corr-a}
\frac{\overline{A}(w,\overline{w})}{(\overline{w}-w)^{1/2}} =
 \frac{\omega}{\sqrt{2}}
\left[-\frac{1}{2}\left\langle{\sigma_1}(w,\overline{w})\right\rangle\right]
\end{equation}
\begin{multline}
\label{eqn:anti-OPE-corr-b'}
 B(w,\overline{w})+\frac{A(w,\overline{w})}{\overline{w}-w}
=\\  \frac{\omega}{\sqrt{2}}(\overline{w}-w)^{1/2}\!
\left(2\partial_{\overline{w}}-i\lambda-\frac{1}{4}\frac{1}{\overline{w}-w}\right)
\left\langle{\sigma_1}(w,\overline{w})\right\rangle
\end{multline}
\begin{multline}
\label{eqn:anti-OPE-corr-c'}
-\frac{A(w,\overline{w})}{(\overline{w}-w)^{5/2}}
=  \frac{\omega}{\sqrt{2}}(w-\overline{w})^{1/2}
\biggl[4{\partial^2_{\overline{w}}} - 4i\lambda{\partial_{\overline{w}}} +
\frac{1}{\overline{w}-w}{\partial_{\overline{w}}}\\ -
\frac{1}{2}i\lambda\frac{1}{w-\overline{w}}
-\frac{1}{16}\frac{1}{(w-\overline{w})^2}\biggr]\!\!
\left\langle{\sigma_1}(w,\overline{w})\right\rangle
\end{multline}

Eqs. \ref{eqn:OPE-corr-a} and \ref{eqn:anti-OPE-corr-a}
imply that $A(w,\overline{w})=-\overline{A}(w,\overline{w})$.
This relation allows us to take the difference between
Eqs. \ref{eqn:OPE-corr-b'} and \ref{eqn:anti-OPE-corr-b'}
to find
\begin{equation}
\label{eqn:time-trans}
\left({\partial_w} + \partial_{\overline{w}}\right)
\left\langle{\sigma_1}(w,\overline{w})\right\rangle=0
\end{equation}
Chatterjee and Zamolodchikov \cite{Chatterjee94}
specialize to the case $w=-\overline{w}=ix$, but this is not
necessary. At no point in the preceding derivation,
leading to Eqs. \ref{eqn:OPE-corr-b'},  \ref{eqn:OPE-corr-c'},
\ref{eqn:anti-OPE-corr-b'},  \ref{eqn:anti-OPE-corr-c'},
do we need $w=(\overline{w})^*$.
Thus, we can take $w={v_n}{\tau_0}+i{x_a}$ and
$\overline{w}={v_n}{\tau}-i{x_a}$.
Then, $w+\overline{w}={v_n}(\tau+{\tau_0})$,
so that Eq. \ref{eqn:time-trans} states that
the correlation function is time-translation invariant,
as expected. Hence, without loss of generality,
we can set ${\tau_0}=0$ so that
$w=i{x_a}$ and
$\overline{w}={v_n}{\tau}-i{x_a}$.

Since the correlation function is independent
of $w+\overline{w}$, we can rewrite (\ref{eqn:OPE-corr-a})
as an ordinary differential equation
in terms of the scaling variable $y=-i\lambda(w-\overline{w})=
\lambda(2{x_a}+i{v_n}\tau)$:
\begin{equation}
\left[-4{\partial_y^2}+ \left(4-\mbox{$\frac{1}{y}$}\right){\partial_y}+
\left(\mbox{$\frac{1}{2y}-\frac{9}{16}\frac{1}{y^2}$}\right)
 \right] \langle {\sigma_1}(y)\rangle = 0
\end{equation}
We warn the reader that there is a typo in Ref. \onlinecite{Chatterjee94},
where $1/y$ appears instead of $1/2y$ in the third term.

If we let $\langle {\sigma_1}(y)\rangle\equiv y^{3/8}\,f(y)$, then
\begin{equation}
yf'' + (1-y)f'-\mbox{$\frac{1}{2}$}f = 0
\end{equation}
This is Kummer's equation, $yf'' + (b-y)f'-af = 0$,
with $a=\frac{1}{2}$ and $b=1$. It has
two linearly independent solutions.
The confluent hypergeometric function (or Kummer's function)
of the first kind, denoted by ${_1F_1}(a,b,y)$ or $M(a,b,y)$,
diverges exponentially for large $y$. The other solution
is the confluent hypergeometric
function (or Kummer's function) of the second kind, denoted
by $U(a,b,y)$. When $\text{Re}(y)>0$, it has the integral representation
\begin{equation}
\label{eqn:hyper-integral-rep}
U(a,b,y)=\frac{1}{\Gamma(a)}{\int_0^\infty}\!dt\,
e^{-ty}\, t^{a-1}\, (1+t)^{b-a-1}
\end{equation}
The confluent hypergeometric functions are singular at
$y=0,\infty$ and can be elsewhere in
terms of formal power series.

From the integral representation (\ref{eqn:hyper-integral-rep}),
we see that $U(a,b,y)$ decays as $1/y^{a}$ for large $y$. 
Hence, this is the appropriate solution for
$\langle {\sigma_1}(y)\rangle$,
leading to $\langle {\sigma_1}(y)\rangle =
({\rm const.})\,\times\,y^{3/8}\,\, U(\frac{1}{2},1,y)$.
The constant is fixed \cite{Chatterjee94} by matching
to the lowest order perturbative calculation so that it
agrees for small $y$ with, for example,
Ref. \onlinecite{Overbosch07,Rosenow07b}:
\begin{equation}
\label{eqn:sigma-crossover}
\langle {\sigma_1}(w,\overline{w})\rangle = 
\lambda^{1/2}\,2^{1/4}\,y^{3/8}\,\,
U\!\left(\mbox{$\frac{1}{2}$},1,y\right)
\end{equation}
where $y=-i\lambda(w-\overline{w})$.


\section{Continuous Neumann and Continuous
Dirichlet Fixed Points}

The crossover discussed in this paper can also be
formulated in terms of an Ising model with a defect
line, rather than a boundary\cite{Fendley09}.
According to this alternative
mapping, the top and bottom edges become the
left- and right-moving sectors of the Ising model.
The `defect line' is the middle of the device, $x=0$,
where we have put a bulk quasiparticle. There are quotation
marks in the previous sentence because it may not be clear
that there is actually a defect at $x=0$ until
one considers the fact that the correlation function
between the tunneling operators at the two point
contacts $\langle {T_a^{}}(0)  {T_b^\dagger}(\tau)\rangle$,
translates to the correlation between two
Ising spin operators, one to the left and one
to the right of the defect,
$\langle\sigma(x,0)\,\sigma(x',\tau)\rangle$
with $xx'<0$.
(Note that we have formed non-chiral Ising
spins in a different way than we did in
Section \ref{sec:mapping}; they are formed from
chiral fields on opposite edges.) With a quasiparticle
in the bulk (which we have not yet coupled to the edge),
this correlation function vanishes. Thus,
the defining feature of the defect line is that this
correlation function vanishes but correlation functions
of spin fields all of which are to the right of the defect
line or all of which are to the left of the defect line
are precisely the same as if there were no defect,
as if there were no bulk quasiparticle.

This defect line doesn't have a simple interpretation
in the classical $2D$ Ising model, but it does in
the $(1+1)$-D transverse field Ising model, where it
corresponds to the quantum Hamiltonian \cite{Oshikawa97}:
\begin{equation}
\label{eqn:Oshikawa-Ham}
H = -h{\sum_{n\neq 0}} {\sigma^x_n}
-J{\sum_{n\neq 0}} {\sigma^z_{n-1}}{\sigma^z_n}
-{J'} \sigma^z_{-1} {\sigma^x_0}
\end{equation}
with $h=J$ in order to tune to criticality and, for
the moment, we specialize to $J'=J$.
At the critical point, we can take the continuum limit,
with $x=na$, where $a$ is the lattice spacing.
The Hamiltonian (\ref{eqn:Oshikawa-Ham})
has the aforementioned property,
$\langle\sigma(x,\tau)\,\sigma(x',\tau')\rangle=0$
for $xx'<0$,
since it is obtained from the usual critical
$(1+1)$-D transverse field Ising model by
performing a duality transformation
\begin{eqnarray}
{\sigma_n^z} \rightarrow {\mu_m^z} &=& \prod_{0\leq m\leq n}
{\sigma^x_n}\cr
{\sigma_n^x} \rightarrow {\mu_n^x} &=&
{\sigma^z_n} \sigma^z_{n+1} 
\end{eqnarray}
on only half of the chain, $n\geq 0$. Thus,
the correlation function
$\langle\sigma(x,\tau)\,\sigma(x',\tau')\rangle=0$
with $xx'<0$
is equal to $\langle\mu(x,\tau)\,\sigma(x',\tau')\rangle=0$
in the ordinary critical $(1+1)$-D transverse field Ising model.
The latter correlation function, between an order and
a disorder field, vanishes.

In fact, this property holds all along the fixed line
obtained by varying $J'$, which was dubbed the
continuous Neumann line in Ref. \onlinecite{Oshikawa97}.
As discussed in Ref. \onlinecite{Fendley09},
varying $J'$ corresponds to
pinching the quantum Hall bar so that interedge
backscattering of Majorana fermions (but not of
charged quasiparticles) can occur at $x=0$. When
Majorana fermions are also allowed to tunnel from
the edges to the bulk $e/4$ quasiparticle at $x=0$,
the system flows from the continuous Neumann line
to the continuous Dirichlet line, at which the bulk 
$e/4$ quasiparticle at $x=0$ has been absorbed by
the edge(s). The continuous Dirichlet line is described by the
critical transverse field Ising model with one bond weakened/strengthened:
\begin{equation}
\label{eqn:cont-Dirichlet}
H = -h{\sum_{n\neq 0}} {\sigma^x_n}
-J{\sum_{n\neq 0}} {\sigma^z_{n-1}}{\sigma^z_n}
-\tilde{J} \sigma^z_{-1} {\sigma^z_0}
\end{equation}

In this paper, we consider the special case
in which there is no constriction at $x=0$, so that
there is no interedge backscattering of Majorana
fermions at $x=0$. Furthermore, the bulk
$e/4$ quasiparticle at $x=0$ is close to only one edge.
Thus, our flow is a special case of the
continuous Neumann to continuous Dirichlet flow
discussed in Refs. \onlinecite{Oshikawa97,Fendley09}.
The combination of Ising correlation functions which enters
the formula (\ref{eqn:P-Ising-mapping}) for the current is simply
$\langle\sigma(x,\tau)\,\sigma(-x,\tau')\rangle=0$.
The nice feature of this formulation is that it refers only
to explicitly single-valued correlation functions in a perturbed
version of \ref{eqn:Oshikawa-Ham}.

\acknowledgements
We thank Paul Fendley for numerous discussions.
We thank A. Stern for a discussion
after one of us (C.N.) gave a talk reporting the present work
at the KITP Conference on Low-Dimensional
Electron Systems in February 2009. We also thank
B. Rosenow, B. Halperin, S. Simon, and A. Stern
for showing us their preprint \cite{Rosenow09} prior to publication.
They find similar results to ours, but by a different method.
The relation of their results to ours can be seen using
the following identity between confluent hypergeometric functions
and modified Bessel functions:
$$
U(n+1/2,1,y) = e^{y/2} {K_n}(y/2)/\sqrt{\pi}
$$


\end{document}